\documentstyle[eqsecnum,aps,preprint]{revtex}

\def\be{\begin{equation}}
\def\ee{\end{equation}}
\def\bea{\begin{eqnarray}}
\def\eea{\end{eqnarray}}

\begin{document}

\preprint{SOGANG-HEP 278/00}

\date{\today}

\title{Higher dimensional flat embeddings of black strings \\
in (2+1) dimensions}

\author{Soon-Tae Hong\footnote{electronic address:sthong@ccs.sogang.ac.kr}, 
Won Tae Kim\footnote{electronic address:wtkim@ccs.sogang.ac.kr},
John J. Oh\footnote{electronic address:john5@string.sogang.ac.kr}, and 
Young-Jai Park\footnote{electronic address:yjpark@ccs.sogang.ac.kr}}

\address{Department of Physics and Basic Science Research Institute,\\
Sogang University, C.P.O. Box 1142, Seoul 100-611, Korea} 

\maketitle

\begin{abstract}
We obtain (3+1) and (3+2) dimensional global flat embeddings of (2+1) 
uncharged and charged black strings, respectively.  In particular, the 
charged black string, which is the dual solution of the 
Banados-Teitelboim-Zanelli black holes, is shown to be embedded in the 
same global embedding Minkowski space structure as that of the (2+1) 
charged de Sitter black hole solution.  
\end{abstract}

\vspace{1cm}
\hspace{1.2 cm} 
%Keywords: black string, BTZ, global flat embedding
PACS number(s): 04.70.Dy, 04.62.+v, 04.20.Jb

\newpage
%%%%%%%%%%%%%%%%%%%%%%%
%\section{Introduction}

%%%%%%%%%%%%%%%%%%%%%%%

It has been well-known that a thermal Hawking effect on a curved 
manifold~\cite{hawk75} can be looked at as an Unruh effect~\cite{unr} 
in a higher flat dimensional space time.  Recently, the isometric embeddings 
of the Reissner-Nordstr\"om (RN) black hole~\cite{deser99} and M2-, D3-, 
M5-branes~\cite{gibbons00} into flat manifolds with two times have been 
studied to yield some insight into the global aspects of the space-time 
geometries in the context of brane physics.  Following the global 
embedding Minkowski space (GEMS) approach~\cite{kasner,rosen65}, several 
authors~\cite{deser99,deser97,des,kps99,kps00} recently have shown that this 
approach could yield a unified derivation of temperature for various curved 
manifolds such as rotating Banados-Teitelboim-Zanelli 
(BTZ)~\cite{btz1,cal,kps,mann93}, Schwarzschild~\cite{sch} together 
with its anti-de Sitter (AdS) extension, RN~\cite{rn} 
and RN-AdS~\cite{kps99}.  Historically, the higher dimensional global flat 
embeddings of the black hole solutions are subjects of great interest to 
mathematicians as well as physicists.  In differential geometry, it has been 
well-known that the four dimensional Schwarzschild metric is not embedded 
in $R^{5}$~\cite{spivak75}.  Very recently, Deser and Levin firstly obtained 
(5+1) dimensional global flat embeddings of (3+1) Schwarzschild black hole 
solution~\cite{deser99,deser97}.  Moreover, the (3+1) dimensional RN-AdS, RN 
and Schwarzschild-AdS black holes are shown to be embedded in (5+2) dimensional 
GEMS manifolds~\cite{kps99}.  

On the other hand, the static, rotating and charged (2+1) dimensional BTZ AdS 
black holes are shown to have (2+2), (2+2) and (3+3) GEMS 
structures~\cite{deser99,deser97,kps00}, while the static, rotating and 
charged (2+1) dimensional de Sitter (dS) black holes are shown to have (3+1), 
(3+1) and 
(3+2) GEMS structures, respectively~\cite{kps00}.  Recently, the (2+1) 
dimensional BTZ black hole~\cite{btz1} has attracted much attention as a 
useful model for realistic black hole physics~\cite{cal} in spite of the fact 
that the space-time curvature is constant.  Moreover, it has been discovered 
the novel aspects that the thermodynamics of higher dimensional black holes 
can often be interpreted in terms of the BTZ solution~\cite{high}, and a 
slightly modified solution of the BTZ black hole yields an solution to the 
string theory, so-called the black string~\cite{horowitz93}. Here one notes 
that this black string solution is in fact only a solution to the lowest order 
$\beta$-function equation to receive quantum corrections~\cite{callan85}.  
It is now interesting to study the geometry of (2+1) dimensional black 
strings, which are the dual solutions of the BTZ black holes, and their 
thermodynamics through further investigation, together with those of the BTZ 
solutions.

In this Brief Report we will analyze Hawking and Unruh effects of the
(2+1) dimensional black strings in terms of the GEMS approach. First, 
we briefly recapitulate the known rotating (2+1) BTZ black
hole and the corresponding charged black string.
Next, we will treat the global higher dimensional flat embeddings of 
the (2+1) uncharged and charged black strings.  In particular, we will
show that the charged black string is embedded in (3+2) GEMS
structure. 

%%%%%%%%%%%%%%%%%%%%%%%%%%%%%%%%%%%%%%%%%%%%%%%%%%
%\section{BTZ Black Hole and Black String Dualities}
%%%%%%%%%%%%%%%%%%%%%%%%%%%%%%%%%%%%%%%%%%%%%%%%%%

Now we briefly summarize the duality properties of the black 
strings given in Ref.~\cite{horowitz93}.  These black string solutions are 
dual ones for the well known (2+1) dimensional rotating BTZ black holes which 
are described by the 3-metric, antisymmetric tensor and dilaton as follows 
\bea
ds^2&=&-N^2dt^2+N^{-2}dr^{2}+r^{2}(d{\phi}+N^{\phi}dt)^{2},\nonumber\\
B_{\phi t}&=&\frac{r^{2}}{l},\nonumber\\
\Phi&=&0.
\label{3metricr} 
\eea 
Here the lapse and shift functions are given as
\be
N^{2}=-M+ \frac{r^2}{l^2} + \frac{J^2}{4r^2} ,~~
N^{\phi}=-\frac{J}{2r^2}, 
\ee 
respectively.  Note that for the nonextremal case there exist
two horizons $r_{\pm}(J)$ satisfying the following
equations,
\be
0=-M+ \frac{r_{\pm}^2}{l^2}+\frac{J^2}{4r_{\pm}^2},
\ee
respectively.  Without solving these equations explicitly, one can then 
rewrite the mass $M$ and angular momentum $J$ in terms of
these outer and inner horizons as follows
\be
M=\frac{r_{+}^2 + r_{-}^2}{l^2},~~ J=\frac{2r_{+}r_{-}}{l}.
\ee
Furthermore these relations can be used to yield the lapse and shift 
functions of the forms 
\be
N^2= \frac{(r^2 - r_{+}^2)(r^2 - r_{-}^2)}{r^2 l^2},~~
N^{\phi}=-\frac{r_{+}r_{-}}{r^2 l},
\ee
respectively. Here one notes that this BTZ space originates from
AdS one via the geodesic identification $\phi=\phi+2\pi$.

On the other hand, one can obtain dual solution \cite{buscher} 
through the transformation 
\bea
g_{\phi\phi}^{d}&=&g_{\phi\phi}^{-1},~~g_{\phi\alpha}^{d}=B_{\phi\alpha}g_{\phi\phi}^{-1},~~
g_{\alpha\beta}^{d}=g_{\alpha\beta}
-(g_{\phi\alpha}g_{\phi\beta}-B_{\phi\alpha}B_{\phi\beta})g_{\phi\phi}^{-1},
\nonumber\\
B_{\phi\alpha}^{d}&=&g_{\phi\alpha}g_{\phi\phi}^{-1},~~
B_{\alpha\beta}^{d}=B_{\alpha\beta}-2g_{\phi[\alpha}B_{\beta]\phi}
g_{\phi\phi}^{-1},\nonumber\\
\Phi^{d}&=&\Phi-\frac{1}{2}\ln g_{\phi\phi},
\label{dualtrfm}
\eea
where $\alpha$, $\beta$ run over all directions except $\phi$, and 
$B_{\phi t}=r^{2}/l$.  The corresponding dual solution is then given as
\be
ds_{d}^{2}=-N^{2}dt^{2}+N^{-2}dr^{2}+\frac{1}{r^{2}}(d\phi_{d} + N_{d}^{\phi}dt)^{2},
\label{3metricd} 
\ee 
where the shift function, antisymmetric tensor and dilaton are defined 
as
\bea
N_{d}^{\phi}&=&B_{\phi t}=\frac{r^2}{l},\nonumber\\
B_{\phi t}^{d}&=&N^{\phi}=-\frac{J}{2r^{2}},\nonumber\\
\Phi^{d}&=&-\ln r, 
\eea 
respectively.  Here one notes that $\phi_{d}/r^{2}$ is periodic such that 
$\phi_{d}/r^{2}=\phi_{d}/r^{2}+2\pi$ and, as expected, the Hawking 
temperature~\cite{hawk75} and entropy are dual invariant
\bea
T_{H}^{d}&=&T_{H}=\frac{r (r_{+}^{2}-r_{-}^{2})}{2\pi 
r_{+}l(r^{2}-r_{+}^{2})^{1/2}(r^{2}-r_{-}^{2})^{1/2}},
\label{th} \\
S_{BH}^{d}&=&S_{BH}=2\pi r_{+}.
\label{sbh}
\eea

Now one can diagonalize the above dual metric with the coordinate 
transformation
\be
t=\frac{l(\hat{x}-\hat{t})}{(r_{+}^{2}-r_{-}^{2})^{1/2}},~
\phi_{d}=\frac{r_{+}^{2}\hat{t}-r_{-}^{2}\hat{x}}{(r_{+}^{2}-r_{-}^{2})^{1/2}},~
r^{2}=l\hat{r}.
\ee
The (2+1) dimensional charged black string solution~\cite{hh} is then given 
by the 3-metric, antisymmetric tensor and non-vanishing dilaton as follows 
\bea
d\hat{s}^{2}&=&-\left(1-\frac{{\cal M}}{\hat{r}}\right)d\hat{t}^{2}
+\left(1-\frac{{\cal Q}^{2}/{\cal M}}{\hat{r}}\right)d\hat{x}^{2}
+\left(1-\frac{{\cal M}}{\hat{r}}\right)^{-1}
 \left(1-\frac{{\cal Q}^{2}/{\cal M}}{\hat{r}}\right)^{-1}\frac{l^{2}
 d\hat{r}^{2}}{4\hat{r}^{2}},\nonumber\\
\hat{B}_{\hat{x}\hat{t}} &=& \frac{\cal Q}{\hat{r}},\nonumber\\
\hat{\Phi}&=&-\frac{1}{2}\ln\hat{r}l,
\label{dshat2bp}
\eea
where ${\cal M}=r_{+}^{2}/l$ and ${\cal Q}=J/2$.  Here note 
that the charge of the black string is linearly proportional to the angular 
momentum of the rotating BTZ black hole. 

%%%%%%%%%%%%%%%%%%%%%%%%%%%%%%%%%%%%%%%%%%%%%%%%%%
%\section{ GEMS Structures of (2+1) Black Strings}
%%%%%%%%%%%%%%%%%%%%%%%%%%%%%%%%%%%%%%%%%%%%%%%%%%

Next, in order to construct the GEMS structures of the black strings, we 
consider the uncharged black string 3-metric
\be
d\hat{s}^{2}=-\left(1-\frac{{\cal M}}{\hat{r}}\right)d\hat{t}^{2}
+d\hat{x}^{2}
+\left(1-\frac{{\cal M}}{\hat{r}}\right)^{-1}\frac{l^{2}d\hat{r}^{2}}
{4\hat{r}^{2}}.
\ee

After algebraic manipulation, we obtain the (3+1) dimensional GEMS
$d\hat{s}^{2}=-(dz^{0})^2+(dz^{1})^2+(dz^{2})^2+(dz^{3})^{2}$ given by the 
coordinate transformations with only one additional space-like dimension 
$z^{3}$, 
\bea
z^{0}&=&\l\left(1-\frac{\hat{r}_{H}}{\hat{r}}\right)^{1/2}
\sinh\frac{\hat{t}}{l}, \nonumber \\
z^{1}&=&\l\left(1-\frac{\hat{r}_{H}}{\hat{r}}\right)^{1/2}
\cosh\frac{\hat{t}}{l}, \nonumber \\
z^{2}&=&\hat{x}, \nonumber \\
z^{3}&=&-l\left(1+\frac{\hat{r}_{H}}{\hat{r}}\right)^{1/2}
+l\ln \left(\hat{r}^{1/2}+(\hat{r}+\hat{r}_{H})^{1/2}\right),
\label{zzz0}
\eea
where $\hat{r}_{H}={\cal M}$.  We will later show how to construct the 
GEMS structure of the more general case of  charged black string, which can 
yield the above result (\ref{zzz0}) in the uncharged limit.  Here note that 
the embedding dimension of this solution is the same as that of the (2+1) 
uncharged dS case~\cite{jac84}, dual version of BTZ black hole.  For the 
trajectories, which follow the Killing vector $\xi=\partial_{\hat{t}}$ on the 
uncharged black string manifold described by $(\hat{t},\hat{r},\hat{x})$, one 
can obtain constant 3-acceleration 
\be
\hat{a}=\frac{\hat{r}_{H}}{l\hat{r}\left(1-\frac{\hat{r}_{H}}{\hat{r}}\right)^{1/2}}.
\ee

We see how the uncharged black string solution yields a finite Unruh area 
due to the periodic identification of $\phi_{d}/(l\hat{r})$ mod $2\pi$.  
The Rindler horizon condition $(z^1)^2- (z^0)^2 = 0$ implies 
$\hat{r}=\hat{r}_H$ and the remaining embedding constraints yield 
$z^{2}=\hat{x}$ and $z^{3}=f(r)$ where $f(r)$ can be read off from 
Eq. (\ref{zzz0}).  The area of the Rindler horizon~\cite{gib77} yields the 
entropy of the uncharged black string
\bea
\hat{S}&=&\int^{\pi (l\hat{r}_{H})^{1/2}}_{- \pi (l\hat{r}_{H})^{1/2}}{\rm d}z^{2}{\rm d}z^{3}\delta(z^{3}-f(r))\nonumber\\
       &=&\int^{\pi (l\hat{r}_{H})^{1/2}}_{- \pi (l\hat{r}_{H})^{1/2}}
       {\rm d}z^{2}=2\pi \hat{r}_{H}.
\label{entropy0}
\eea

Now we construct the GEMS structure of the charged black string.  In order to 
obtain the $d\hat{t}^{2}$ term in Eq. (\ref{dshat2bp}) we make an ansatz 
of two coordinates $(z^{0}, z^{1})$ in Eq. (\ref{dsch}) to yield 
\be
-(dz^0)^2+(dz^1)^2=-\left(1-\frac{\hat{r}_{H}}{\hat{r}}\right)d\hat{t}^{2}
+\frac{1}{4}\left(\frac{l^{2}\hat{r}_{H}}
{\hat{r}_{H}-\hat{r}_{-}}\right)\left(\frac{\hat{r}_{H}}
{\hat{r}^{2}}\right)^{2}\left(1-\frac{\hat{r}_{H}}{\hat{r}}\right)^{-1}
d\hat{r}^{2}
\label{dz01}
\ee
where $\hat{r}_{H}={\cal M}$, and $\hat{r}_{-}={\cal Q}^{2}/{\cal M}$.  
Similarly, the $d\hat{x}^{2}$ term in 
Eq. (\ref{dshat2bp}) can be constructed by exploiting an ansatz of two 
coordinates $(z^{2}, z^{3})$ in Eq. (\ref{dsch}) as follows 
\be
-(dz^2)^2+(dz^3)^2=\left(1-\frac{\hat{r}_{-}}{\hat{r}}\right)d\hat{x}^{2}
-\frac{1}{4}\left(\frac{l^{2}\hat{r}_{-}}
{\hat{r}_{H}-\hat{r}_{-}}\right)
\left(\frac{\hat{r}_{-}}{\hat{r}^{2}}\right)^{2}
\left(1-\frac{\hat{r}_{-}}{\hat{r}}\right)^{-1}d\hat{r}^{2}.
\label{dz23}
\ee
Since the combination of Eqs. (\ref{dz01}) and (\ref{dz23}) yields
\bea
& &-(dz^0)^2+(dz^1)^2-(dz^2)^2+(dz^3)^2
=-\left(1-\frac{\hat{r}_{H}}{\hat{r}}\right)d\hat{t}^{2}
   +\left(1-\frac{\hat{r}_{-}}{\hat{r}}\right)d\hat{x}^{2}
\nonumber\\
& &+\left(1-\frac{\hat{r}_{H}}{\hat{r}}\right)^{-1}
    \left(1-\frac{\hat{r}_{-}}{\hat{r}}\right)^{-1}
 \frac{l^{2}d\hat{r}^{2}}{4\hat{r}^{4}}
 \left[\hat{r}_{H}^{2}+\hat{r}_{H}\hat{r}_{-}+\hat{r}_{-}^{2}
-\frac{\hat{r}_{H}\hat{r}_{-}(\hat{r}_{H}+\hat{r}_{-})}{\hat{r}}\right]
\nonumber\\
& &=d\hat{s}^{2}-\frac{l^{2}d\hat{r}^{2}}{4\hat{r}^{3}}
(\hat{r}+\hat{r}_{H}+\hat{r}_{-})
\equiv d\hat{s}^{2}-(dz^{4})^{2},
\label{dz0123}
\eea
we obtain the desired (3+2) GEMS $d\hat{s}^{2}=-(dz^{0})^2+(dz^{1})^2
-(dz^{2})^2+(dz^{3})^{2}+(dz^{4})^{2}$ for the charged black string~\cite{hh} 
given by the coordinate transformations, 
\bea
z^{0}&=&\left(\frac{l^{2}\hat{r}_{H}}{\hat{r}_{H}-\hat{r}_{-}}\right)^{1/2}
\left(1-\frac{\hat{r}_{H}}{\hat{r}}\right)^{1/2}
\sinh \left(\frac{\hat{r}_{H}-\hat{r}_{-}}{l^{2}\hat{r}_{H}}\right)\hat{t}, 
\nonumber \\
z^{1}&=&\left(\frac{l^{2}\hat{r}_{H}}{\hat{r}_{H}-\hat{r}_{-}}\right)^{1/2}
\left(1-\frac{\hat{r}_{H}}{\hat{r}}\right)^{1/2}
\cosh \left(\frac{\hat{r}_{H}-\hat{r}_{-}}{l^{2}\hat{r}_{H}}\right)\hat{t}, 
\nonumber \\
z^{2}&=&\left(\frac{l^{2}\hat{r}_{-}}{\hat{r}_{H}-\hat{r}_{-}}\right)^{1/2}
\left(1-\frac{\hat{r}_{-}}{\hat{r}}\right)^{1/2}
\cosh \left(\frac{\hat{r}_{H}-\hat{r}_{-}}{l^{2}\hat{r}_{-}}\right)\hat{x}, 
\nonumber \\
z^{3}&=&\left(\frac{l^{2}\hat{r}_{-}}{\hat{r}_{H}-\hat{r}_{-}}\right)^{1/2}
\left(1-\frac{\hat{r}_{-}}{\hat{r}}\right)^{1/2}
\sinh \left(\frac{\hat{r}_{H}-\hat{r}_{-}}{l^{2}\hat{r}_{-}}\right)\hat{x}, 
\nonumber \\
z^{4}&=&-l\left(1+\frac{\hat{r}_{H}+\hat{r}_{-}}{\hat{r}}\right)^{1/2}
+l\ln \left(\hat{r}^{1/2}+\left(\hat{r}+\hat{r}_{H}+\hat{r}_{-}
\right)^{1/2}\right).
\label{dsch}
\eea
This equivalence between the (3+2) GEMS metric associated with Eq. 
(\ref{dsch}) and the original curved space metric in Eq. (\ref{dshat2bp}) is 
the very definition of isometric embedding, mathematically developed by several 
authors~\cite{gibbons00,nash54}.  Here one can also easily check that, in the 
uncharged limit ${\cal Q}\rightarrow 0$ 
(or $\hat{r}_{-}\rightarrow 0$), the above 
coordinate transformations $(z^{0},z^{1},z^{4})$ for the case of 
${\cal Q}\neq 0$ are exactly reduced to $(z^{0},z^{1},z^{3})$ for the case of 
${\cal Q}=0$ in the previous ones (\ref{zzz0}), while $(z^{2},z^{3})$ with 
${\cal Q}\neq 0$ merge into one positive definite coordinate $z^{2}(=\hat{x})$ 
with ${\cal Q}=0$ since in this uncharged limit $-(dz^2)^2+(dz^3)^2\rightarrow 
d\hat{x}^{2}$ in Eq. (\ref{dz23}).  Note that this isometric embedding 
(\ref{dsch}) is equivalent to the charged dS case~\cite{jac84}, which is the 
dual version of charged BTZ black hole.  For the trajectories, which follow the
Killing vector $\xi=\partial_{\hat{t}}$ on the charged black string manifold 
described by $(\hat{t},\hat{r},\hat{x})$, one can obtain constant 
3-acceleration 
\be
\hat{a}=\frac{\hat{r}_{H}\left(1-\frac{\hat{r}_{-}}{\hat{r}}\right)^{1/2}}{l\hat{r}
\left(1-\frac{\hat{r}_{H}}{\hat{r}}\right)^{1/2}}.
\ee

Now we see how the charged black string solution yields a finite Unruh 
area due to the periodic identification of $\phi_{d}/(l\hat{r})$ mod 
$2\pi$.  The Rindler horizon condition $(z^1)^2- (z^0)^2 = 0$ implies 
$\hat{r}=\hat{r}_{H}$ and the remaining embedding constraints yield 
\bea
z^{4}&=&f(r),\nonumber\\
(z^2)^2 - (z^3)^2 &=&(l{\cal Q}/{\cal M})^{2},
\label{constem}
\eea
where $f(r)$ can be read off from Eq. (\ref{dsch}).  Here one notes that such 
local and isometric embedding spaces have the global topology that cover only 
the area outside the horizon $\hat{r}_{H}$ with the above embedding constraints 
(\ref{constem}) and cannot be extended to the entire 
solution~\cite{rosen65,deser97}.

With this global topology of the isometric embedding in mind, we can obtain 
the area of the Rindler horizon described as  
\[
\int {\rm d}z^{2}{\rm d}z^{3}{\rm d}z^{4}\delta(z^{4}-f(r))\delta\left([(z^2)^2-(z^3)^2]^{1/2}
-\frac{l{\cal Q}}{{\cal M}}\right)
\]
which, after performing trivial integrations over $z^4$, 
yields the desired entropy of the charged black string
\bea
\hat{S}&=&\int^{(l{\cal Q}/{\cal M})\sinh (\pi \hat{r}_H/(l{\cal Q}^{2}
/{\cal M})^{1/2})}_{- (l{\cal Q}/{\cal M})\sinh (\pi \hat{r}_H/(l{\cal Q}^{2}
/{\cal M})^{1/2})}{\rm d}z^{3}\int^{[(z^3)^2+(l{\cal Q}/{\cal M})^2]^{1/2}}_{0}
{\rm d}z^{2}\delta\left([(z^2)^2-(z^3)^2]^{1/2}-\frac{l{\cal Q}}{{\cal M}}
\right)\nonumber\\
     &=&\int^{(l{\cal Q}/{\cal M})\sinh (\pi \hat{r}_H/(l{\cal Q}^{2}
     /{\cal M})^{1/2})}_{-(l{\cal Q}/{\cal M})\sinh (\pi \hat{r}_H
     /(l{\cal Q}^{2}/{\cal M})^{1/2})}{\rm d}z^{3}\frac{l{\cal Q}/{\cal M}}
     {[(l{\cal Q}/{\cal M})^2 +(z^3)^2]^{1/2}}=2\pi \hat{r}_{H}. 
\eea

%%%%%%%%%%%%%%%%%%%%%%%%%%%%%%%%%%%%%%%%%%%%%%%
%\section{Conclusion}
%%%%%%%%%%%%%%%%%%%%%%%%%%%%%%%%%%%%%%%%%%%%%%%

In conclusion, we have investigated the higher dimensional global flat embeddings of (2+1) 
uncharged and charged black strings.  These black strings are shown to be 
embedded in the (3+1) and (3+2)-dimensions for the uncharged and 
charged black strings, respectively.  Here it is worthwhile noting that 
both the uncharged and charged black strings have the same GEMS structures as 
those of uncharged and charged de Sitter black holes, which are dual 
solutions of the corresponding BTZ anti-de Sitter black holes.  Moreover, 
through further investigation, it will be interesting to study the topology 
of boundary surface of black holes associated with the nontrivial higher genus 
and the corresponding Yamabe invariant~\cite{anderson97,gibbons99cqg}, in the 
framework of the GEMS structure.

\vspace{1cm}

\acknowledgements{
We would like to thank Y.-W. Kim, J.S. Park and M.-I. Park for helpful 
discussions.  We acknowledge financial support from the Ministry of 
Education, BK21 Project No. D-1099 and the Korea Research Foundation, 
Project No. KRF-2000-015-DP0070.}

\end{document}